\documentclass[aps,epsfig,prl,twocolumn]{revtex4}
\usepackage[dvips]{graphicx}
\usepackage[dvips]{color}
\usepackage{subfigure}
\usepackage{epsfig}
\usepackage{dcolumn}
\usepackage{bm}
\usepackage{amsmath}%
\usepackage{amssymb} 
\newcommand\beq{\begin{equation}}
\newcommand\bear{\begin{eqnarray}}
\newcommand\eeq{\end{equation}}
\newcommand\eear{\end{eqnarray}}

\begin{document}

\title{Structural Relaxation of a Gel Modeled by Three Body Interactions}
\author{Shibu Saw$^{1}$, Niels  L. Ellegaard$^{1}$, Walter Kob$^{2}$, Srikanth Sastry$^{1}$}
\affiliation{$^{1}$ Theoretical Sciences Unit, Jawaharlal Nehru Centre for Advanced 
Scientific Research, Jakkur Campus, Bangalore 560 064, India.\\
$^{2}$ Laboratoire des Collo\"ides, Verres et Nanomat\'eriaux, UMR5587 CNRS, 
Universit\'e Montpellier 2, 34095 Montpellier, France}

\date{\today}

\begin{abstract}
We report a molecular dynamics simulation study of a model gel
whose interaction potential is obtained by modifying the three body
Stillinger-Weber model potential for silicon. The modification reduces
the average coordination number, and suppresses the liquid-gas phase
coexistence curve. The low density, low temperature equilibrium gel
that can thus form exhibits interesting dynamical behavior, including
compressed exponential relaxation of density correlations. We show
that motion responsible for such relaxation has ballistic character,
and arises from the motion of chain segments in the gel without the
restructuring of the gel network.
\end{abstract}

\pacs{xx}

\maketitle

Gels are low density disordered networks of interacting molecules that
are structurally arrested and capable of sustaining weak
stresses. They are ubiquitous in nature and among man-made materials
and are composed of a diverse range of materials such as polymers,
silica, or colloidal particles. Depending on the life time of the
bonds between the basic units of the network, they can either be
chemical gels, or reversible, physical gels, the latter displaying
complex dynamics. In particular, colloidal gel formers exhibit
intricate dynamic behavior in equilibrium, as well as in nonequilibrium
aging conditions, can form arrested states, and have been the subject
of a considerable number of experimental, theoretical and simulation
studies
\cite{Zacca,Cip04a,bibette,Cipelletti-PRL-2000,Manley-2005-PRL,Bandyopadhyay,soga,Bergenholtz-1999-PRE,ZacPRL2005,bianchi,Sastry-JSM,delgado,cates,Zac04b,KobEuro,KobPRL,hurtado,hurtado2,suarez,pitard}.

One reason for the interest in colloidal gels is that these systems permit access to the glassy state via several mechanisms: Cluster
aggregation \cite{bibette,suarez}, structural arrest in the dense
phase following phase separation \cite{Zac04b,Manley-2005-PRL},
or crossing of a glass transition line from an equilibrium
fluid to an arrested state \cite{Bergenholtz-1999-PRE,KobEuro}.
For the occurrence of this latter scenario it is necessary that
upon cooling the system does not enter the liquid-gas coexistence
region~\cite{Sastry-2000-PRL,ZacPRL2005,bianchi,Sastry-JSM,KobEuro},
i.e. one seeks systems for which the coexistence region is at low
temperatures, $T$, and densities, $\rho$. One possibility to achieve
this is to choose a ``maximum valency'' interaction, in which each
particle can interact only with a restricted (small) number of
particles~\cite{ZacPRL2005,bianchi,Sastry-JSM}.  In the following we will show
that a very simple model involving three body interactions is also
able to generate a coexistence region that is located at low $T$ and
$\rho$ and which therefore allows one to  probe easily the interplay of
phase transformations and dynamics in molecular dynamics simulations in
such systems.

A further intriguing property of colloidal gels
is the fact that their relaxation dynamics can be
compressed~\cite{Cipelletti-PRL-2000,Bandyopadhyay}, i.e. the time
correlation functions decay faster than an exponential, in stark contrast
to structural glasses at higher densities for which one usually finds a
stretched exponential relaxation. The microscopic origin of this fast relaxation
is not well understood, and various mechanisms have been proposed to
explain it~\cite{Cip04a,KobPRL,pitard}. We present analysis that shows that for our model, compressed exponential relaxation arises from the ballistic motion of chain segments in the gel without restructuring the gel network.

The model we consider is a modification of the potential
proposed by Stillinger and Weber (SW) for the description of
silicon~\cite{SW}. Particles interact via a sum of two and three body
interaction terms, $v = v_2(r) + \lambda v_3(r,\theta)$~\cite{swfootnote},
where $r$ denotes interparticle distances and $\theta$ the angle
formed by three particles. $\lambda$ determines the strength of
the three body interaction which depends on the angle $\theta$ via a term proportional to 
$(\cos\theta+\alpha)^{2}$ with $\alpha$ determining the most preferred
angle. Thus by varying $\lambda$ \cite{molinero} and $\alpha$ we can tune the locally
preferred arrangement of the particles.

We have performed constant temperature, volume molecular dynamics (MD) simulations
(using a constraint that conserves kinetic energy) with 4000 particles, using the
method proposed in \cite{W-S1993,Makhov-Lewis,shibu2} to efficiently compute
three body interactions. Gibbs-Ensemble-Monte-Carlo (GEMC) simulations
\cite{Panagiotopoulos-gemc} are performed to obtain liquid-gas coexistence curves
have been performed with $2000$ particles.  All results 
are reported in reduced units for the Stillinger-Weber potential \cite{SW}.

\begin{figure}[th]
\includegraphics[scale=0.30]{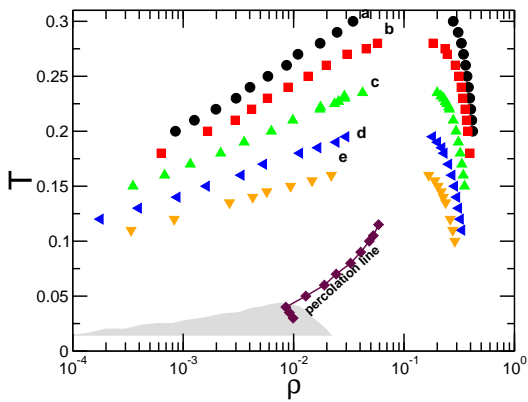}
\caption{
Phase coexistence curves calculated from GEMC simulations for (a)
$\lambda$=21, $\alpha$=1/3, (b) $\lambda$=25, $\alpha$=1/3, (c)
$\lambda$=25, $\alpha$=1/2, (d) $\lambda$=10, $\alpha$=1.00 and (e)
$\lambda$=10, $\alpha$=1.10. With the increase of value $\lambda$ or
$\alpha$ the phase coexistence curve is suppressed.  The shaded area
indicates the expected coexistence region for $\lambda=10.0$ and
$\alpha=1.49$. Also shown is the percolation line for these values of
$\lambda$ and $\alpha$.
}
\label{phase-coex}
\end{figure}

Figure~\ref{phase-coex} shows the coexistence curves obtained for
various combinations of $\lambda$ and $\alpha$. We see that with
increasing $\lambda$ or $\alpha$, liquid-gas phase coexistence gets shifted
to smaller temperature and density ranges, analogous to the observations
in \cite{ZacPRL2005,bianchi,Sastry-JSM}. In the following we will fix
$\lambda=10.0$ and $\alpha=1.49$. For this choice the structure of the
system at low $T$ and $\rho$ is given by quasi-one-dimensional chains
of particles, interconnected by three coordinated junctions. At low $T$
bond breaking becomes extremely difficult and consequently a reliable
estimate of the coexistence curve via GEMC is no longer possible. Based on MD
runs where we observe signatures of phase separation, we indicate in
Fig. \ref{phase-coex} the region where we expect phase separation (shaded
area). Also included in the graph is the percolation line which indicates
the density and temperature range (to the lower right of the percolation
line) where we may expect gel-like structural arrested states. (The bend
in the percolation line at low temperature is due to phase separation \cite{shibu2}.) In
the following we will study the relaxation dynamics of the system for
$\rho=0.06$ and from Fig.~\ref{phase-coex} it is clear that at this
density phase separation will not play a role.

\begin{figure}[thb]
\includegraphics[scale=0.40]{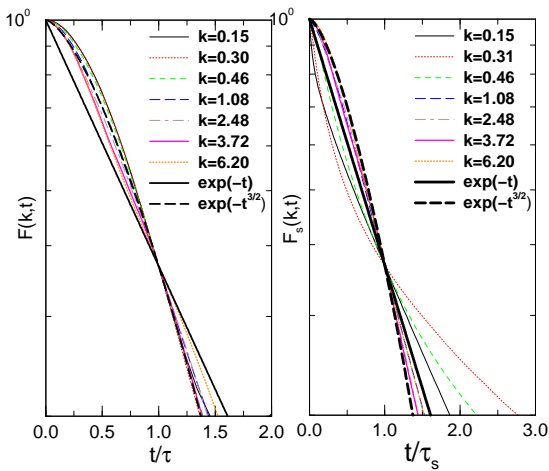}
\caption{
The collective intermediate scattering function $F(k,t)$ (left panel)
and self intermediate scattering function $F_s(k,t)$ (right panel)
at $T = 0.03$, for a range of wave-vectors $k$.
}
\label{FktFskt}
\end{figure}

In order to characterize the relaxation dynamics of the system we
consider the normalized collective, and self intermediate scattering
functions, defined as $F(k,T) = S(k)^{-1} \sum_{j,l} \langle \exp[-i
  {\bf k} \cdot ({\bf r}_j(t) - {\bf r}_l(0))] \rangle$, where $S(k)$
is the static structure factor, and $F_s(k,T) = N^{-1} \sum_{j}
\langle \exp[-i {\bf k}\cdot ({\bf r}_j(t) - {\bf r}_j(0))] \rangle$,
respectively.  The time dependence of these correlators is shown, in
semi-log plots, in Fig.~\ref{FktFskt}, for the low temperature $T =
0.03$, i.e. well below the percolation line, and various values of the
wave-vector $k$. At this $T$ the relaxation dynamics of the system is
already very sluggish and hence we deal here indeed with a
glass-forming system (see Fig. \ref{tau-area} in which one observes a strong change of the relaxation times with temperature, and Ref.~\cite{shibu2}
for a detailed discussion).  Since we look for compressed exponentials,
we plot the data as a function of $t/\tau$ and $t/\tau_s$, where
$\tau$ and $\tau_s$ are the relaxation times defined by requiring that
the correlator has decayed to $e^{-1}$ of its initial value. We see
that the two time correlation functions display remarkably different
behavior: For intermediate wave-vectors $F(k,t)$ curves downward and
can be fitted well by the Kohlrausch-Williams-Watts (KWW) function
$A\exp(-(t/\tau)^{\beta})$ with $\beta \approx 3/2$, the so-called
compressed exponential (CE). [If $k$ is very small or very large the
  decay is even faster, i.e. $\beta \geq 1.5$ (not shown)]. Such a
behavior ($\beta \approx 3/2$) has been observed in experiments of
slowly relaxing gels \cite{Cipelletti-PRL-2000,Bandyopadhyay}, and
analyzed theoretically using a stress relaxation model
\cite{pitard}. On the other hand $F_s(k,t)$ shows, like most other
glass-forming systems, a stretched exponential, i.e. $\beta \leq 1.0$
for small and intermediate wave-vector, and a compressed exponential
at large wave-vectors. Such a behavior has also been observed in
\cite{KobPRL}, and interpreted as arising from the averaging of
ballistic motion of particles which form a part of chain segments of
varying lengths in the disordered percolating network.

\begin{figure}[b]
\includegraphics[scale=0.30]{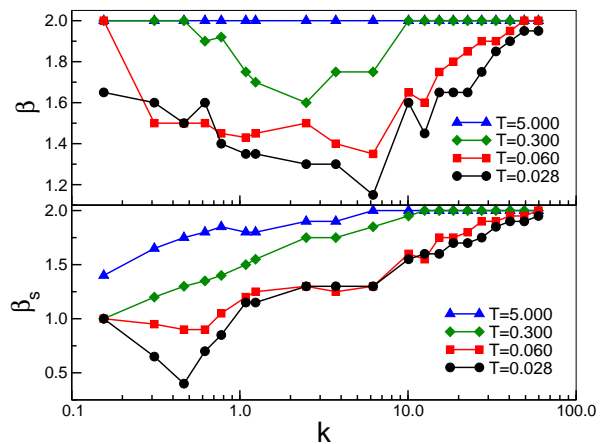}
\caption{
Wave-vector dependence of the KWW exponent $\beta$ for $F(k,t)$ and $\beta_s$ for $F_s(k,t)$ 
for four different temperatures.}
\label{beta}
\end{figure}

To analyze further the nature of the relaxation, we determined the
KWW-exponent $\beta$ by fitting the correlators to a KWW-function. Since
the time correlators exhibit different regimes of decay, it is necessary
to choose a meaningful and consistent procedure for obtaining $\beta$
and we choose to fit the curve in the time window in which the correlator
is between $0.9$ and $0.1$. This choice avoids  the
(trivial) ballistic regime at very short times and instead focuses
on the relaxation regime seen at intermediate times.  The wave-vector
dependence of the so obtained KWW-exponents is shown in Fig.~\ref{beta}
for four different temperatures. The $\beta$ values for $F(k,t)$, shown
in the top panel, are seen to be always $2.0$ for $T=5.0$, indicating
that ballistic motion dominates the decay (as discussed in the context of gels in \cite{KobPRL}) at all wavelengths. While this
is expected for large $k$, we note that the behavior at small $k$ is a
result of the low density of our system which leads to significant decay
of collective density fluctuations even on large wavelengths through
non-diffusive motion of particles. If $T$ is decreased $\beta$ shows
a minimum at intermediate values of $k$ and the width of this minimum
broadens with decreasing $T$, suggesting that on intermediate length
scales the decay mechanism is distinct. The typical values of $\beta$
in this minimum are around $1.3-1.6$, {\it i.e.} similar to the values
that have been found in the experimental systems or in the theoretical
calculations.

The $\beta_s$ values for $F_s(k,t)$, shown in the bottom panel, are found,
for the highest temperature $T = 5.0$, to change from $2.0$ at large $k$
towards $1.0$ at small $k$, which we interpret as the expected crossover
from ballistic to diffusive decay. Again, for low temperatures, we find
superposed on this overall trend an intermediate regime, in which the
dynamics becomes ``stretched'', {\it i.e.}, the decay of the self motion
becomes slower than exponential. Thus from this figure we can conclude
that the self and collective density correlation functions exhibit
complex behavior, that is non-trivial, and different from dense fluids.

\begin{figure}[th]
\includegraphics[scale=0.40]{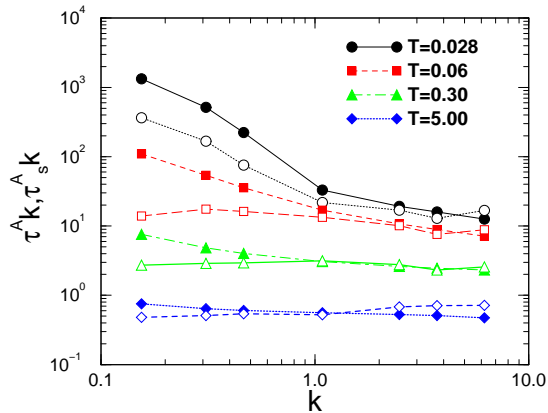}
\caption{
The wave-vector dependence of relaxation time from the area of $F(k,t)$
and $F_s(k,t)$ for temperature $0.028$, $0.06$ , $0.30$ and $5.00$ at density
$0.06$. The filled symbols are for $\tau_s$ and opaque symbols indicates
$\tau$. }
\label{tau-area}
\end{figure}

The relaxation times, $\tau^{\rm A}(k)$ and $\tau_s^{\rm A}(k)$, obtained
from calculating the area under $F(k,t)$ and $F_s(k,t)$, are shown in
Fig. \ref{tau-area} as a function of $k$ for different temperatures. Since
for  ballistic motion one expects the relaxation time to be proportional
to $k^{-1}$, we show $\tau^{\rm A}(k)$ and $\tau_s^{\rm A}(k)$ multiplied by
$k$. From the figure we recognize that at high and intermediate $T$
this scaling gives indeed horizontal line, showing that the motion
can be interpreted as ballistic. Furthermore we see that the self and
collective relaxation times track each other for all $k$.  At low $T$ and
small $k$ the curves are no longer horizontal, indicating that there is 
a significant non-ballistic component. 
As we shall see below, at these low temperatures, 
$F(k,t)$ has a significant long time relaxation that is clearly distinguishable 
from an intermediate relaxation process which we shall identify with CE behavior. 
Furthermore there is a strong decoupling at low $k$ in that $\tau_s^{\rm A}$ exceeds $\tau^{\rm
A}$ by a large factor.

\begin{figure}[th]
\includegraphics[scale=0.40]{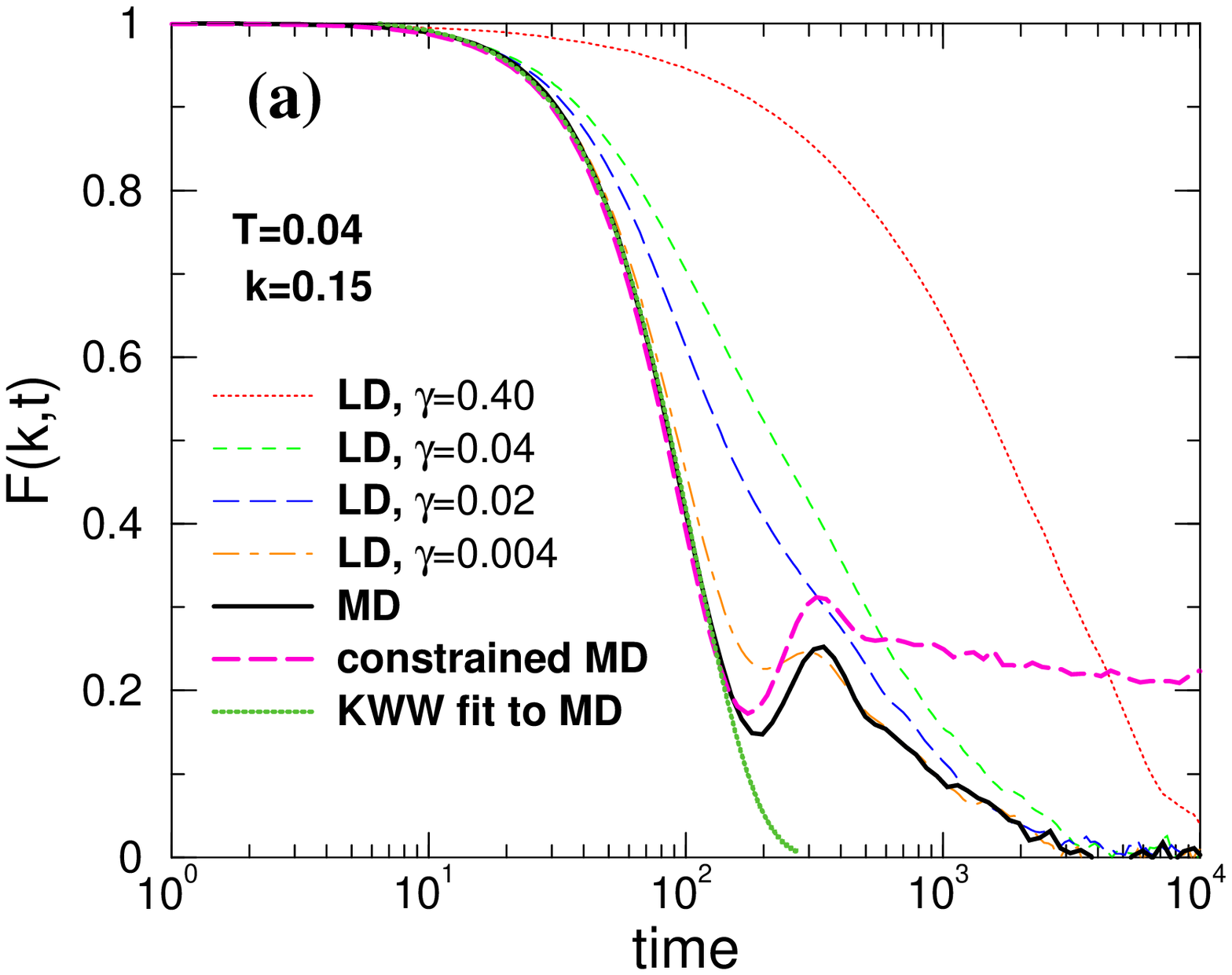}
\includegraphics[scale=0.40]{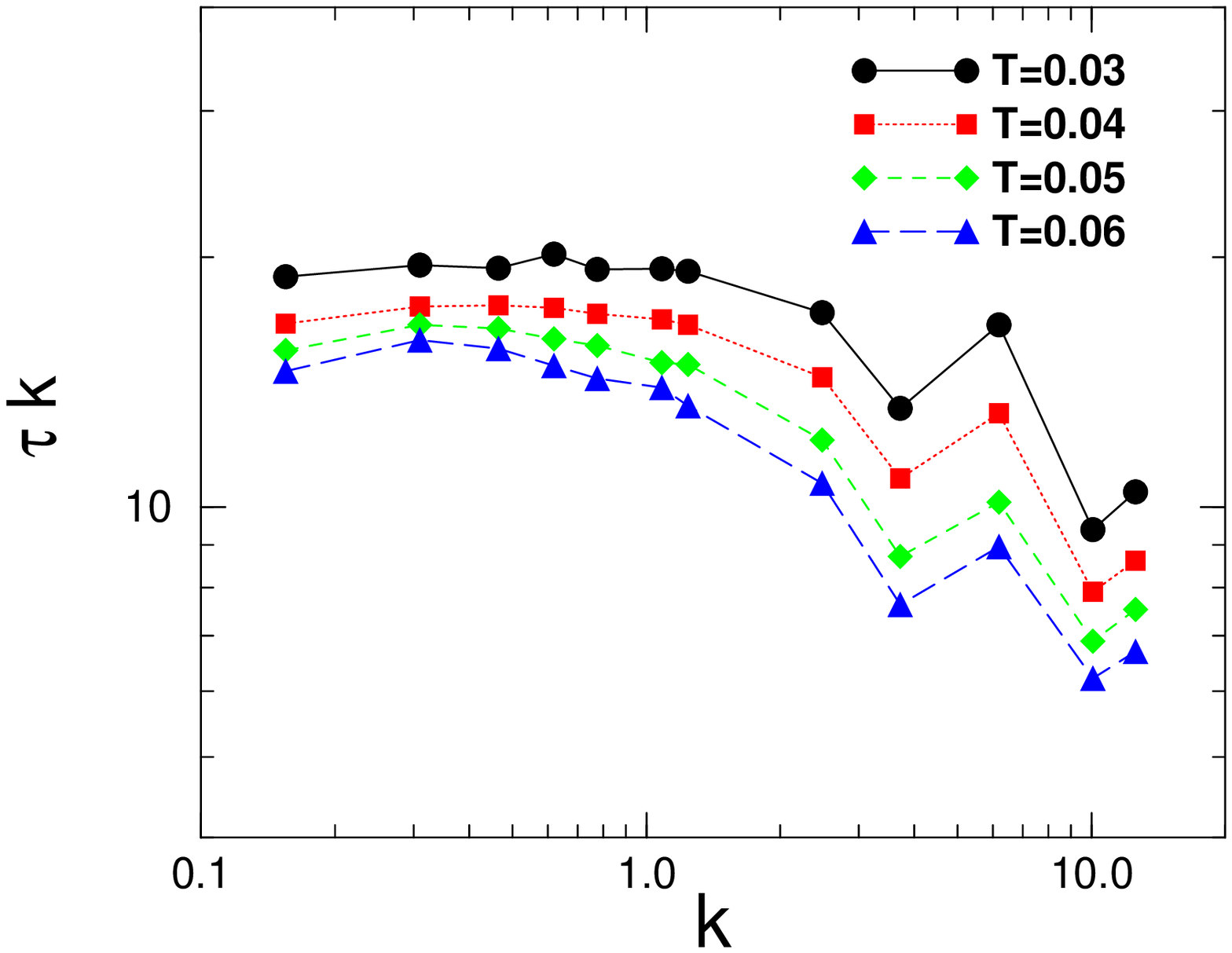}
\caption{(a)
The $F(k,t)$ from molecular dynamics (MD), and Langevin dynamics (LD)
for different damping constants $\gamma$.  Also shown is $F(k,t)$ from
constrained MD simulation described in the text, and the KWW fit to the MD curve for intermediate times. (b)  $\tau$k $vs$ wave-vector, $k$ for
a range of  temperatures,  showing that at low $k$, $\tau \sim 1/k$.}
\label{tauk-LD}
\end{figure}

We now investigate the nature of the compressed exponential relaxation
of $F(k,t)$ (shown for $T = 0.04, k = 0.15$ in Figure 5(a)), which we
observe on intermediate time scales. At low temperatures, CE
relaxation accounts for a substantial part of the decay of $F(k,t)$
for a wide range of intermediate $k$ values. Since at these low
temperatures the average life time of the bonds is longer than the
decay time of $F(k,t)$ in the CE regime \cite{shibu2}, we expect the CE to be associated with the
floppy dynamics of chain segments in the transient gel network,
without network restructuring playing any role. In order to analyze
the motions that are relevant, we therefore compare the $F(k,T)$
obtained in MD simulations to: (a) MD simulations with the imposition
of a constraint that prevents bond breaking and formation of new
bonds. This is accomplished by identifying bonded neighbors in the
initial configuration we consider, and adding a suitably parametrized
barrier potential of gaussian form to the two body part of the S-W
potential.  (b) Langevin dynamics simulations, to study the role of
microscopic dynamics. For the Langevin dynamics we have used a
predictor-corrector integrator \cite{Beard-JCP-2003}, and the damping
coefficient is tuned to span the range from very small damping,
$\gamma = 0.004$ (corresponding to MD) to strong damping,
$\gamma=0.4$. In Fig. \ref{tauk-LD} (a) we show $F(k,t)$ from
these different simulations.

Comparing MD with the constrained MD results, we see that the regime of
CE dynamics is essentially unaltered by the imposition of the constraint
not to break or form bonds. This shows clearly that the CE dynamics arises
from the dynamics of the non-restructuring gel network. However, at longer
times the relaxation dynamics of the constrained MD is essentially frozen,
indicating that long time relaxation in the MD, cleanly separated from
the compressed exponential decay, arises from network restructuring, a
result that is also confirmed by the time dependence of the mean squared
displacement of the particles (not shown)~\cite{shibu2}.

In Fig. \ref{tauk-LD} (b) we show $\tau \times k$ where $\tau(k)$ is
obtained by fitting $F(k,t)$ by a KWW function in the intermediate
time window displaying CE relaxation, as shown in Figure 5 (a) (note
that these $\tau$ values are different from those shown in
Fig. \ref{tau-area} which are obtained from the area under
$F(k,t)/S(k)$). The near constant value of $\tau \times k$ for small $k$ corroborates the
{\it ballistic} origin of the compressed exponential relaxation, consistent
with predictions \cite{Cipelletti-PRL-2000,pitard}. 

For the Langevin dynamics we see that for small and intermediate
values of the damping coefficient $\gamma$, the correlator tracks the one from the MD and thus a CE will
be observed. However, for large damping the shape of the curve is very
different from the one of the MD and no CE is seen anymore. Thus we see
that the dissipative dynamics, relevant for example for real colloidal
gels, will not show a CE dynamics. Therefore we can conclude that the
CE seen in those systems is likely due to the aging dynamics.

In conclusion, we have proposed a model system which allows the
simulation and study of gel forming fluids under equilibrium conditions,
by suppressing the liquid-gas phase coexistence curve to an arbitrarily
small temperature and density window. At low densities and temperatures
the structural and dynamical features show many similarities to the one
of experimental systems. In particular we find an intricate behavior
of the density correlation functions, including compressed exponential
relaxation of the collective intermediate scattering function with
a compressing exponent that depends on temperature and wave-vector
considered. The motion responsible for the compressed relaxation is
found to have ballistic character and to arise due to the motion of
chain segments in the gel without the restructuring of the gel network.

We thank D. Weitz, L. Cipelletti, S. Ciliberto and F. Sciortino for
fruitful discussions. We thank Indo-French Centre for the Promotion of
Advanced Research - IFCPAR for financial support and CCMS, JNCASR for
computational facilities.

\end{document}